\documentclass[conference]{IEEEtran}
\IEEEoverridecommandlockouts
\usepackage{amsmath,amssymb,amsfonts}
\usepackage{algorithmic}
\usepackage{graphicx}
\usepackage{textcomp}
\usepackage{xcolor}
\def\BibTeX{{\rm B\kern-.05em{\sc i\kern-.025em b}\kern-.08em
    T\kern-.1667em\lower.7ex\hbox{E}\kern-.125emX}}
\usepackage[belowskip=-1pt,aboveskip=0pt]{caption}
\usepackage{subcaption}
\usepackage{gensymb}
\usepackage{float}
\usepackage{array}
\usepackage{physics}
\usepackage{optidef}
\usepackage{algorithm}
\usepackage{algorithmic}
\usepackage{bm}

\usepackage{hyperref}
\usepackage{xurl}
\hypersetup{
  colorlinks=true, 
  linkcolor=blue,   
  citecolor=blue,  
  urlcolor=blue     
}
\usepackage{booktabs} 

\usepackage[sort&compress,numbers]{natbib} 

\begin{document}
\title{High-Speed Voltage Control in Active Distribution Systems with Smart Inverter Coordination and DRL\\
\thanks{This work was supported in part by the National Science Foundation (NSF) under the grant ECCS-2145063.}
}
\author{Mohammad Golgol,~\IEEEmembership{Student Member,~IEEE}, and Anamitra Pal,~\IEEEmembership{Senior Member,~IEEE}\\
\IEEEauthorblockA{\textit{School of Electrical, Computer and Energy Engineering} \\
\textit{Arizona State University, Tempe, AZ, USA}\\
mgolgol@asu.edu, anamitra.pal@asu.edu}}

\maketitle

\begin{abstract}
The increasing penetration of renewable energy resources in distribution systems necessitates high-speed monitoring and control of voltage for ensuring reliable system operation. 
However, existing voltage control algorithms often make simplifying assumptions in their formulation, such as \emph{real-time availability} 
of smart meter measurements (for monitoring), or \emph{real-time knowledge} of \textit{every} power injection information (for control).
This paper leverages the recent advances made in high-speed 
state estimation for real-time unobservable distribution systems to formulate a deep reinforcement learning (DRL)-based control algorithm that utilizes the state estimates \textit{alone} to control the voltage of the entire system.
The results obtained for a modified (renewable-rich) IEEE 34-node distribution feeder indicate that the proposed approach excels in monitoring and controlling voltage of active distribution systems.
\end{abstract}



\vspace{-2mm}

\section{Introduction}
The imperative to reduce carbon emissions has spurred the integration of renewable energy resources into modern power systems.
However, the substantial incorporation of renewable energy, 
particularly at the distribution-level in the form of rooftop solar photovoltaic (PV), poses serious challenges to reliable system operation. 
High penetration of PVs have caused elevated voltages and voltage fluctuations in active distribution systems \cite{amanipoor2022v,singh2021event}.
Furthermore, conventional voltage control devices for the distribution system such as on-load tap changers (OLTCs), distribution voltage regulators (DVRs), and/or capacitor banks (CBs) 
\cite{suresh2022coordinated}
are not designed for frequent use (e.g., multiple times in a day).
This implies that relying on them primarily for voltage control in renewable-rich distribution grids will cause increased wear-and-tear, and eventually accelerate loss-of-life of these equipment.   

Due to the above-mentioned reasons,
current research is focused 
on exploiting the capability of modern smart inverters in providing reactive power (Q) support through Volt-VAr (VV) coordination \cite{li2022cybersecurity}. 
This can be done through  \textit{optimization-based} approaches or \textit{data-driven} approaches.
The former involves optimizing VV settings using deterministic or stochastic methods
\cite{tang2020distributed,zhang2021three,dalal2023improving}.
However, their reliance on model accuracy as well as lower speed of online operation have been identified as possible shortcomings.
As such, data-driven approaches involving machine learning (ML), particularly 
deep reinforcement learning (DRL), are gaining considerable attention. 
Ref. \cite{zhang2020deep} cast the VV optimization problem within a deep Q-network (DQN) framework to overcome modeling limitations.
Refs. \cite{cao2021data,wang2021multi,cao2023physics} used the soft-actor critic algorithm, while \cite{zhang2021ddpg,liu2021online} employed the deep deterministic policy gradient (DDPG) to solve the voltage control problem.

However, one or more of the following issues were identified with the DRL-based
control approaches 
developed in \cite{zhang2020deep,
cao2021data,wang2021multi,cao2023physics,
zhang2021ddpg,liu2021online}.
Most of them
assumed access to \textit{full system information} such as
power consumption, solar power generation, and voltage, \textit{in real-time}.
In practice, such data is not readily available. For instance, 
smart meters measure voltage and \emph{average} power at intervals of at least 15 minutes \cite{lang2021data}.
More importantly, these measurements
are not reported to the utility until \textit{after a few hours} \cite{azimian2021time}.
This means that they cannot be used directly for performing
distribution system state estimation (DSSE), which was the basis for 
these control architectures.
Furthermore,
these approaches assumed that conventional distribution voltage control devices 
(OLTCs, DVRs, CBs),
are designed for frequent operation, which is not the case in reality.


The control approach proposed in this paper 
advances the state-of-the-art in the following ways:
(1) It sidesteps the need for full system information in real-time by relying solely on
a state 
estimator that uses measurements
\textit{at the feeder-head alone} to perform high-speed DSSE. 
(2) Its training and performance verification
are done using this state estimator.
(3) It
only changes settings of the smart inverters using remote signals
for mitigating voltage violations; i.e., it does not change settings of the conventional control devices.
(4) It causes minimal active power curtailment. 
The results obtained using the IEEE 34-node \textit{unbalanced} distribution feeder proves that the proposed innovative, infrastructure-light approach enables the monitoring and management of active distribution systems that have high penetration of rooftop solar PV. 



\vspace{-1mm}
\section{ML-based Time Synchronized DSSE}
\vspace{-1mm}
In a modern distribution system, sensors are usually placed at the root node (e.g., SCADA, $\mu$PMU) or the leaf nodes (e.g., smart meters, solar meters).
Consequently, 
the main trunk of the system remains concealed from power utilities from monitoring and control perspectives. 
Moreover, the high latency period of smart meters prevents their data from being reported
in real-time.
This limited real-time visibility lowers
the effectiveness of conventional approaches for DSSE \cite{azimian2021time}.

To overcome this challenge and achieve comprehensive situational awareness with minimal sensor coverage, an ML-based high-speed state estimator was developed in \cite{azimian2022state}.
It used Bayesian inference to establish a mapping relationship between $\mu$PMU measurements and state variables (voltage magnitude and angle) of all network nodes.
Historical data from slow-timescale, high-latency sensors were used for training, while fast-timescale data was used for testing.
In this paper, we employ this ML-based state estimator to perform end-to-end high-speed DSSE using $\mu$PMUs placed \textit{only at the feeder-head}.
The DSSE outputs are used to (a) train the proposed DRL-based control, and (b) test the performance of the control.
More details about the offline training and online implementation of this state estimator can be found in \cite{azimian2023analytical,azimian2023book}.

\section{DRL-based Inverter Control} \label{modeling}

\subsection{Modeling voltage control as an optimization problem}\label{AA}
We start by formulating the voltage control problem as a constrained optimization problem, as shown below:
\begin{align}
    \min_{Q_{\text{PV}}} J &= \sum_{t=1}^{T} \left( \sum_{n=1}^{N} \lvert V_{[n,t]}-V_{\mathrm{nominal}} \rvert \right) \label{eq:objective}
\end{align} 
subject to:
\vspace{-5mm}
\begin{equation}
    \begin{aligned}
         V_{R}[n,t] \left( \sum_{m=1}^{N} \left( G_{n,m}V_{R}[m,t] -B_{n,m}V_{I}[m,t]\right)\right) \\
         + V_{I}[n,t] \left(\sum_{m=1}^{N} \left( G_{n,m}V_{I}[m,t] +B_{n,m}V_{R}[m,t]\right) \right)\\
         + P[n,t]= 0 \quad \quad \forall n \in N \text{ and } \forall t \in T
    \end{aligned} \label{eq:constraint1}
\end{equation}
\begin{equation}
    \begin{aligned}
    P[n,t] = P_{\text{load}}[n,t]-P_{\text{PV}}[k,t]-P_F[t] \\ \quad \quad \forall n \in N, t \in T, \text{ and } k \in K 
    \end{aligned} \label{eq:constraint2}
\end{equation}
\textcolor{black}{\begin{equation}
    \begin{aligned}
        0 &\le P_{\text{PV}}[k,t] \le P_{\text{rated}}[k] \quad \forall k \in K, t \in T
    \end{aligned} \label{eq:constraint7}
\end{equation}}
\vspace{-5mm}
\begin{equation}
    \begin{aligned}
        V_{I}[n,t] \left( \sum_{m=1}^{N} \left( G_{n,m}V_{R}[m,t] -B_{n,m}V_{I}[m,t]\right)\right) \\
        - V_{R}[n,t] \left(\sum_{m=1}^{N} \left( G_{n,m}V_{I}[m,t] +B_{n,m}V_{R}[m,t]\right) \right)\\
        + Q[n,t]= 0 \quad \quad \forall n \in N \text{ and } \forall t \in T
    \end{aligned}\label{eq:constraint3}
\end{equation} 
\begin{equation}
    \begin{aligned}
        Q[n,t] = Q_{\text{load}}[n,t]-Q_{\text{PV}}[k,t]-Q_F[t] \\ \quad \quad \forall n \in N, t \in T, \text{ and } k \in K
    \end{aligned} \label{eq:constraint4}
\end{equation}
\textcolor{black}{\begin{equation}
    \begin{aligned}
        -Q_{\text{rated}}[k] &\le Q_{\text{PV}}[k,t] \le Q_{\text{rated}}[k] \quad \forall k \in K, t \in T
    \end{aligned} \label{eq:constraint5}
\end{equation}}
\vspace{-5mm}
\begin{equation}
    \begin{aligned}
        V_{\mathrm{min}} &\le V[n,t] \le V_{\text{max}} \quad  \forall n \in N, t \in T
    \end{aligned} \label{eq:constraint6}
\end{equation}

The objective function \eqref{eq:objective}
minimizes the absolute difference between the voltage of node $n$ at time $t$, denoted by $V[n,t]$, and the nominal rated voltage, denoted by $V_{\mathrm{nominal}}$ (set at 1 p.u. by default)
for all $n$ and $t$.
This objective is attained by adjusting the
Q output of smart inverters.
The constraints that must be satisfied 
are defined in \eqref{eq:constraint1}-\eqref{eq:constraint6}, in which 
$V_{R}[n,t]$ and $V_{I}[n,t]$ denote the real and imaginary components of $V[n,t]$, $G_{n,m}$ and $B_{n,m}$ denote the conductance and susceptance between nodes $n$ and $m$, $P_F$ and $Q_F$ denote the active and reactive power drawn from the feeder-head, and 
$K$, $N$, and $T$ denote sets of PV sources, all nodes' indices, and time-steps, respectively.
Equations \eqref{eq:constraint1}-\eqref{eq:constraint4} ensure active and reactive power balance.
\textcolor{black}{In \eqref{eq:constraint5}, $Q_{\text{rated}}[k]$ refers to the maximum available reactive power that could be provided by the $k^{th}$ smart inverter, while $S_{\text{rated}}[k]$ and $P_{\text{rated}}[k]$ are the rated values of apparent and active power 
for that inverter, respectively.}
The desired voltage limits are introduced in \eqref{eq:constraint6}.


In order to solve \eqref{eq:objective}-\eqref{eq:constraint6}
analytically, one needs 
\textit{instantaneous values} of the power injections,
which are not readily available in most distribution systems. 
Furthermore, solving this problem at each time-step can be time-consuming, particularly as the network size increases.
To address these issues, we propose a data-driven approach based on DRL that employs the DDPG algorithm. 
The proposed approach eliminates the need for precise model parameters and allows the agent to adapt to real-time data. Moreover, the trained agent can be used in real-time without significant response delays. In the following sub-section, we detail how this constrained optimization problem can be transformed into a Markov decision process (MDP), which is the prerequisite for solving any problem using DRL.

\subsection{Formulating voltage control as an MDP problem}
An MDP is a mathematical model for decision-making, involving an agent interacting with an environment over discrete time-steps. Its key components are state, action, environment, and  reward.
The state represents the system's different configurations. In our case, we use voltage magnitudes of every node, as it encapsulates all the necessary information required to make an informed decision.
The following equation specifies the system state, $s$, at time $t$:
\vspace{-2mm}
\begin{align}
    s_t = (|V[n,t]|), \quad \forall n \in N \label{eq:state}
\end{align}

Actions are the choices available to the DRL agent. For our problem, they are 
described by the following equations: 
\textcolor{black}{
\begin{align}
   \begin{split}
        Q_{\text{PV}}[k,t] &= a_t^l \cdot Q_{\text{rated}}[k] \quad \text{where }  -1 \le a_t^l \le 1
   \end{split} \label{eq:Q_curtail}
\end{align}}
\vspace{-5mm}
\begin{equation}
        A_t = (a_t^l, \quad \forall l \in L) \text{ at each time } t  \label{eq:action}
\end{equation}

In \eqref{eq:Q_curtail}, each agent gives a specific coefficient $a_t^l$ that determines how much available reactive power, of those smart inverters that are controlled by it, should be utilized for voltage control.
$L$ is the set of all agents, with \eqref{eq:action} specifying the action of all the agents at time $t$.
The environment models the system's dynamics, determining the next state based on the current state-action pair. 
The reward function defines the immediate benefit or cost of taking the action. For our objective of maintaining voltage within desired range with minimum active power curtailment, 
the following reward function is proposed:

\begin{equation}
        r_t = -\left( \lambda\sum_{n\in N} \Lambda(V[n,t]) + \eta\sum_{k\in K} \Gamma\left(Q_{\text{PV}}[k,t]\right) \right)\label{eq:reward}
\end{equation}
\begin{equation}
        \Lambda(V) = \begin{cases}
        \lVert V-V_{\mathrm{nominal}}\rVert^2 & V_{\mathrm{min}} \le V \le V_{\text{max}}\\
        \lvert V-V_{\mathrm{nominal}}\rvert & \mathrm{otherwise}
    \end{cases} \label{eq:volt_barr} \\
\end{equation}
\begin{align}
    \textcolor{black}{
    \Gamma(q_k) = \begin{cases}
        0 & 0 \le \lvert q_k \rvert \le Q_{\text{max}}[k]\\
        \lvert \lvert q_k\rvert -Q_{\text{max}}[k]\rvert & \mathrm{otherwise}
    \end{cases} \label{eq:Q_barr}}\\
    \textcolor{black}{Q_{\text{max}}[k] = \sqrt{S_{\text{rated}}^2 - P_{\text{PV}}^2[k]} \label{eq:Q_without_curtailment}
    }
\end{align} 
where, $\Lambda(\cdot)$ corresponds to the voltage control effort,
$\Gamma(\cdot)$ corresponds to the power curtailment effort, and
\textcolor{black}{\( Q_{\text{max}} \) refers to the maximum reactive power that can be supplied without active power curtailment \cite{photovoltaics2018ieee}.}
As the cardinality of $N$ and $K$ may differ, we introduce $\lambda$ and $\eta$ to balance 
the two efforts.
Fine-tuning these parameters enables us to align the network's behavior with 
operational guidelines.

\subsection{Control execution using DDPG algorithm}
The DDPG algorithm is well-suited for continuous action spaces \cite{lillicrap2015continuous}, which fits the scope of our problem.
It
consists of an actor and a critic network, which are neural networks parameterized by $\theta_\pi$ and $\theta_\vartheta$, respectively, as shown below:
\begin{equation}
    \begin{aligned}
        A_t = \pi(s_t \mid \theta_\pi)
    \end{aligned} \label{Actor_net}
\end{equation} 
\begin{equation}
    \begin{aligned}
         \vartheta(s_t, A_t | \theta_\vartheta) \approx r_t + \gamma \vartheta(s_{t+1}, \pi(s_{t+1} | \theta_\pi') | \theta_\vartheta')
    \end{aligned} \label{critic_net}
\end{equation}

The actor \eqref{Actor_net} maps states to actions based on the policy function, while the critic evaluates the action 
\eqref{critic_net}. The discount factor 
$\gamma \in [0, 1]$,
governs the consideration of future rewards.  Note that to ensure robustness and stability during training, target networks ($\theta_\pi'$ and $\theta_\vartheta'$) are employed for both the actor and the critic, respectively, which are incrementally updated to shadow the learned networks. The training utilizes a deterministic policy gradient for the actor and a temporal difference (TD) error for the critic.

\subsubsection{Offline Stage}

\begin{figure}
\centerline{\includegraphics[width=0.489\textwidth]{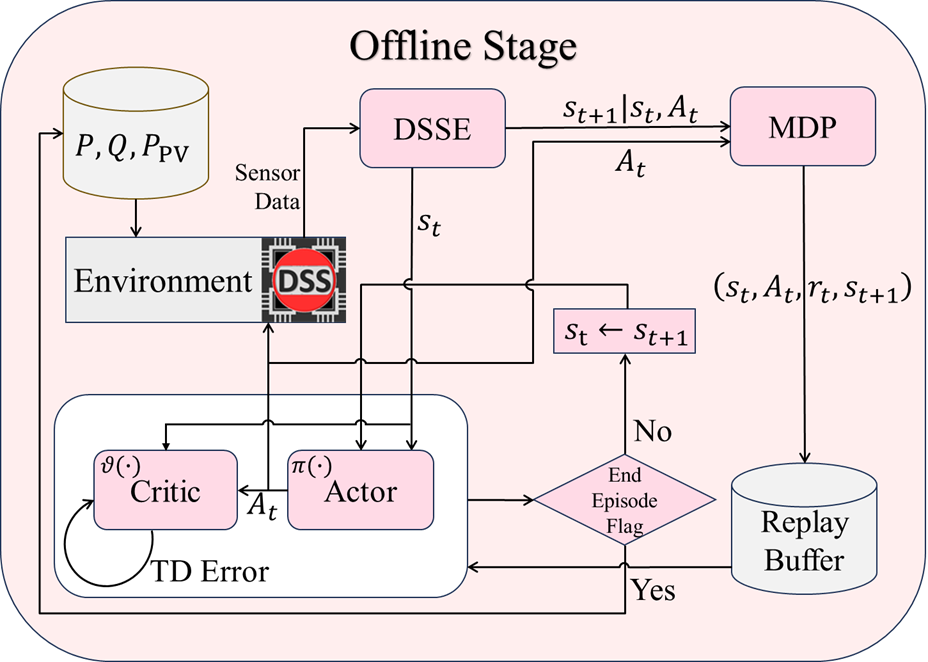}}
\vspace{2mm}
\caption{Offline training process of proposed control}
\label{offline}
\end{figure}

The offline stage, depicted in Fig. \ref{offline}, is the training phase, where the agent is exposed to various power system scenarios, represented by $(P, Q, P_{\mathrm{PV}})$, to learn the optimal control actions.
Each scenario is processed through the environment and DSSE modules to acquire the 
state, $s_t$. Subsequently, the agent determines the action $A_t$ using \eqref{Actor_net}, leading to the new state $s_{t+1}$. The immediate reward $r_t$ is computed given the transition from $s_t$ to $s_{t+1}$. These experiences are stored as transitions in the replay buffer until the end of each episode, and a new scenario is then selected for the next episode. When the replay buffer is filled, a batch of transitions is utilized to adjust the weights of the actor and critic networks iteratively until the TD error converges. 
This is explained mathematically in the following equations:
\begin{equation}
    \begin{aligned}
        \mathrm{TD} = \frac{1}{B} \sum_{i=1}^B \left(  \vartheta(s_i, A_i | \theta_\vartheta) - (r_i + \gamma \vartheta(s_{i+1}, \pi(s_{i+1} | \theta_\pi') | \theta_\vartheta'))\right)
    \end{aligned}\label{TD}
\end{equation}
\begin{equation}
    \begin{aligned}
        \nabla_{\theta_\pi} L \approx \frac{1}{B} \sum_{i=1}^B \nabla_A \vartheta(s,A\mid \theta_\vartheta) \mid_{s_i, A=\pi(s_i)}\nabla_{\theta_\pi}\pi(s\mid \theta_\pi)\mid_{s_i}
    \end{aligned} \label{policy}
\end{equation}
\begin{equation}
    \begin{cases}
        \theta_\vartheta' \longleftarrow \tau \theta_\vartheta + (1-\tau)\theta_\vartheta' \\
        \theta_\pi' \longleftarrow \tau \theta_\pi + (1-\tau)\theta_\pi'
    \end{cases} \label{update_target}
\end{equation}

The update process for the critic network involves minimizing \eqref{TD}, while the actor network undergoes updates through the sampled policy gradient method outlined in \eqref{policy}. Additionally, the parameters of the target networks are updated according to the specifications in \eqref{update_target}. The term $B$ in \eqref{TD}-\eqref{policy}
represents the number of transitions in a batch, while $\tau$ in \eqref{update_target} is the soft update coefficient. 

\begin{figure}
\centerline{\includegraphics[width=0.489\textwidth]{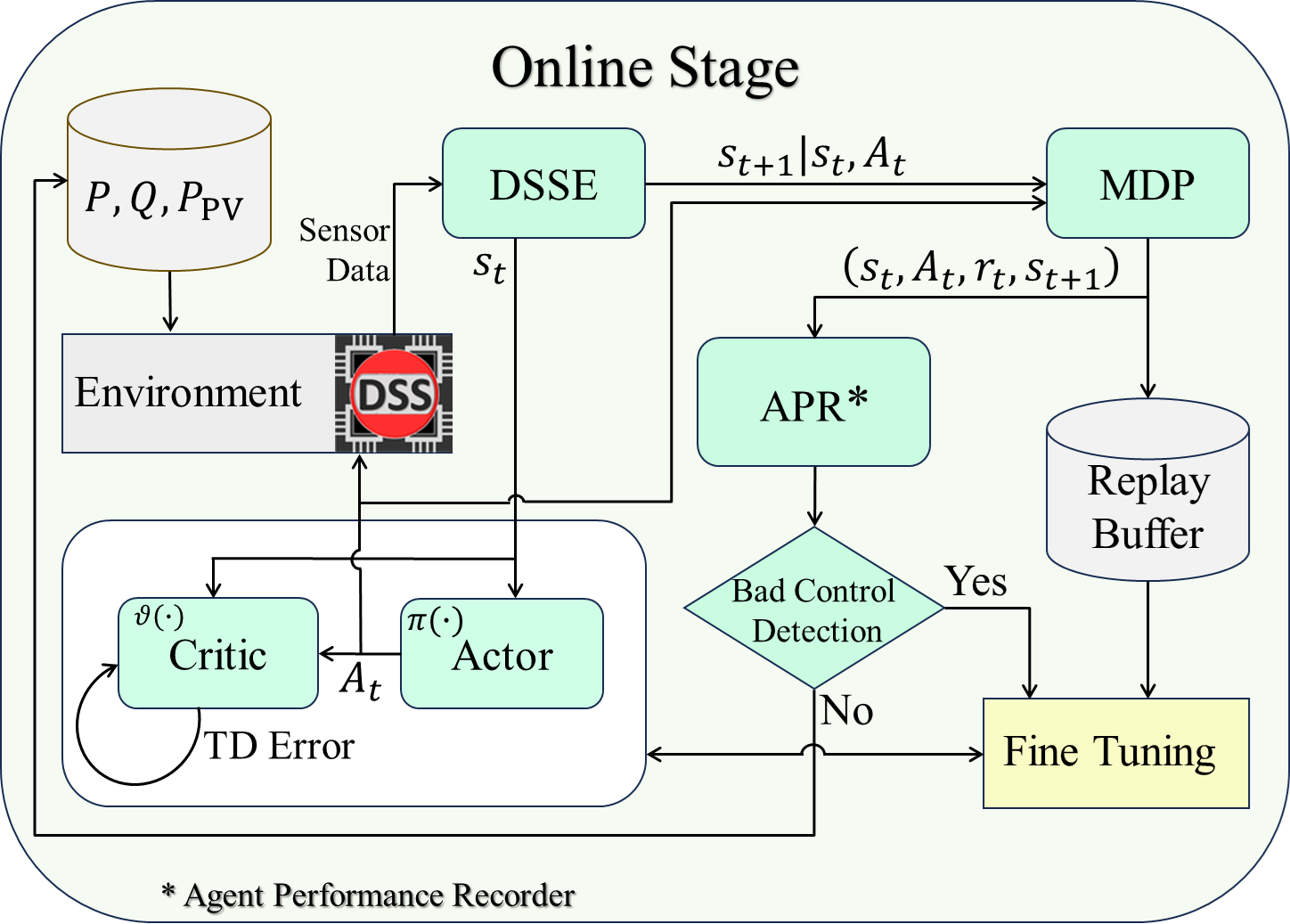}}
\vspace{2mm}
\caption{Online execution of proposed control}
\label{online}
\end{figure}

\subsubsection{Online Stage}

In the online phase, shown in Fig. \ref{online}, the trained agent is subjected to testing.
The DSSE module processes real-time sensor data to estimate the states. The voltage magnitudes obtained at the output of the DSSE module serve as the current state $s_t$ for the agent. 
The agent then determines the appropriate action $A_t$ to adjust the system's state. 
The resulting state $s_{t+1}$ and reward $r_t$
are determined through the DSSE and MDP modules, respectively. The agent performance recorder (APR) continuously monitors the agent's decision-making process (via the DSSE and MDP modules).
If the APR identifies poor control performance, it initiates a fine-tuning process for the agent's parameters, enabling it to adapt to the evolving operational conditions of the system.

\section{Simulation result}
\subsection{Case-study and data preparation}

We conducted simulations on the IEEE 34-node unbalanced feeder
to assess the impact of high PV penetration at the distribution level. 
\textcolor{black}{We added PV systems at all load locations, with PV power generation ranging from $74\%$ to $105\%$ of the respective load.}
As this system lacks historical smart meter data,
\textcolor{black}{we generated diverse scenarios using Pecan Street data \cite{pecandataport}. This dataset contains smart meter load and rooftop PV data from 11 houses. We selected data corresponding to July with a time interval of 12 Noon to 1:00 PM, representing the period of maximum PV power generation.} 
Finally, since the 34-node feeder is at the primary voltage level, we randomly combined household level load and PV profiles from Pecan Street to create aggregated profiles at the primary level. 



\subsection{Parameter settings}
The DNN model in our ML-based DSSE featured a sequential architecture with 5 hidden layers, each consisting of 200 neurons having $\mathrm{ReLU}$ activation function. To prevent overfitting, a $50 \%$ dropout rate and batch normalization were applied after each hidden layer. The DNN, with an input size of 12 (voltage and current phasor values of all three phases at the feeder-head) and output of 270 (voltage magnitudes and angles of all phases of every node), was trained for 100 epochs using the mean squared error loss function.
The optimizer used was Adam and the learning rate was 0.095.

The actor and critic networks in our DDPG algorithm
utilized two hidden layers with 400 and 300 neurons each, a learning rate of 0.001, and a batch size of 64. The activation functions in the critic network were $\mathrm{ReLU}$ for both layers, while it was $\mathrm{ReLU}$ for the first layer and $\mathrm{tanh}$ for the second layer in the actor network. 
\textcolor{black}{The soft update coefficient, denoted by $\tau$, was fixed at 0.001, while the exploration-exploitation parameters, $\mu$ and $\sigma$, were initially set to 0.1 and 0.5, respectively. As the training progressed, $\sigma$ gradually decreased from 0.5 to 0.005.}
The input layer of the actor network had a size of 135, which corresponded to the number of phase voltage magnitudes obtained from DSSE.
The critic network's input layer had a size of $\mathrm{135+1}$, combining state and action information. Both networks had an output layer size of 1. For the reward function, the results were achieved with values of 1 for $\lambda$ and 0.5 for $\eta$. The replay buffer size was set to $10,000$.
\subsection{Performance of ML-based DSSE}
\textcolor{black}{The DSSE results for the modified (renewable-rich) 
IEEE 34-node system with one $\mu$PMU at the feeder-head is shown in Table \ref{DSSE_Res}.}
The mean absolute percentage error (MAPE) metric is used for the magnitudes, while the mean absolute error (MAE) metric is used for the angles.
The results were obtained with 1\% Gaussian noise in the $\mu$PMU data, and confirm the ability of ML to overcome the real-time unobservability issue in distribution systems.
This state estimator is now used to train and test the proposed control.


\begin{table}[H]
\centering
\caption{DSSE results for modified IEEE 34-node system}
\label{DSSE_Res}
\begin{tabular}{lccc}
\toprule 
& Phase A & Phase B & Phase C \\
\midrule 
Magnitude MAPE (\%) & 0.0804 & 0.0957 & 0.0228 \\
Phase MAE ($\degree$) & 0.0248 & 0.0228 & 0.0212 \\
\bottomrule 
\end{tabular}
\end{table}
\vspace{-1.5mm}


\subsection{Assessment of voltage regulation performance}

\begin{figure}[htbp]
\centerline{\includegraphics[width=0.489\textwidth]{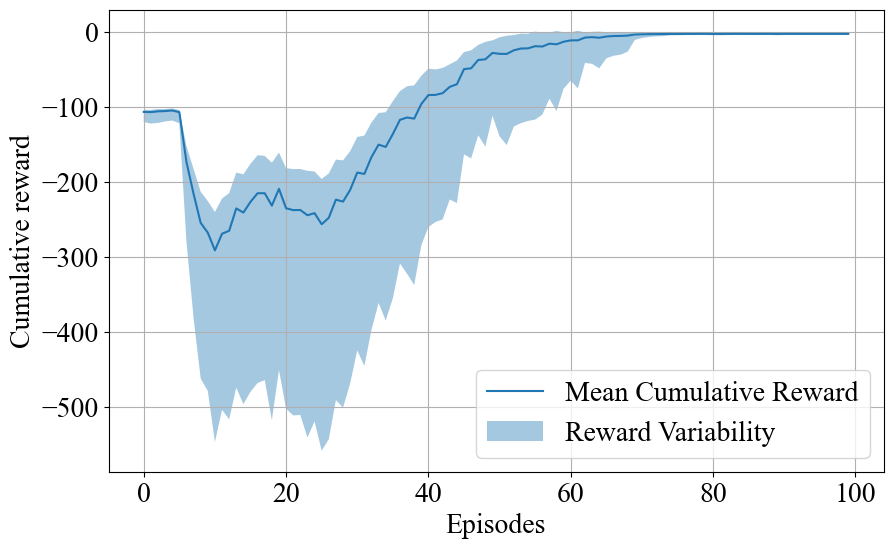}}
\caption{Evolution of cumulative reward  over $100$ training runs}
\label{reward_trajectory}
\end{figure}

Fig. \ref{reward_trajectory} shows the average cumulative reward across $100$ training runs, with the solid line representing the mean reward at each episode and the shaded area indicating the variability in the performance of the agent (captured via standard deviation of the reward).
The initial decline in the reward is due to early exploration and learning.
Over the course of 100 episodes, we observe an overall upward trend in the reward, suggesting that the agent is effectively learning and improving its policy.
The fluctuations in the reward trajectory are due to the exploratory nature of DRL and the stochastic elements within the environment.
The diminishing width of the shaded region over time reflects a reduction in performance variance, pointing to the agent's increasing stability and the algorithm's convergence towards a consistent policy.

\begin{figure}[htbp]
\centerline{\includegraphics[width=0.489\textwidth]{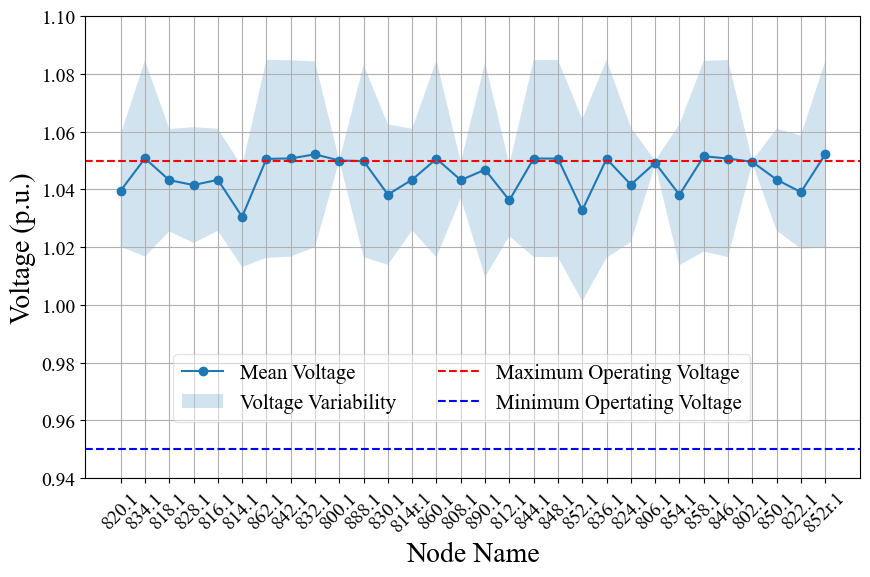}}
\caption{Phase A voltage profile at unity power factor operation
}
\label{volt_without_coor}
\end{figure}

Fig. \ref{volt_without_coor} shows the Phase A voltage magnitude results obtained in the absence of control (i.e., inverters operating at unity power factor) across 10,000 scenarios.
It can be observed from the figure that
the mean voltage (solid blue line) and its variance (shaded area) frequently exceed the upper voltage limit (red dotted line). 
The several instances of over-voltage in the different nodes indicate the potential risks and challenges associated with managing voltage in active distribution grids.
Similar observations were made for the other phases as well.

\begin{figure}[htbp]
\centerline{\includegraphics[width=0.489\textwidth]{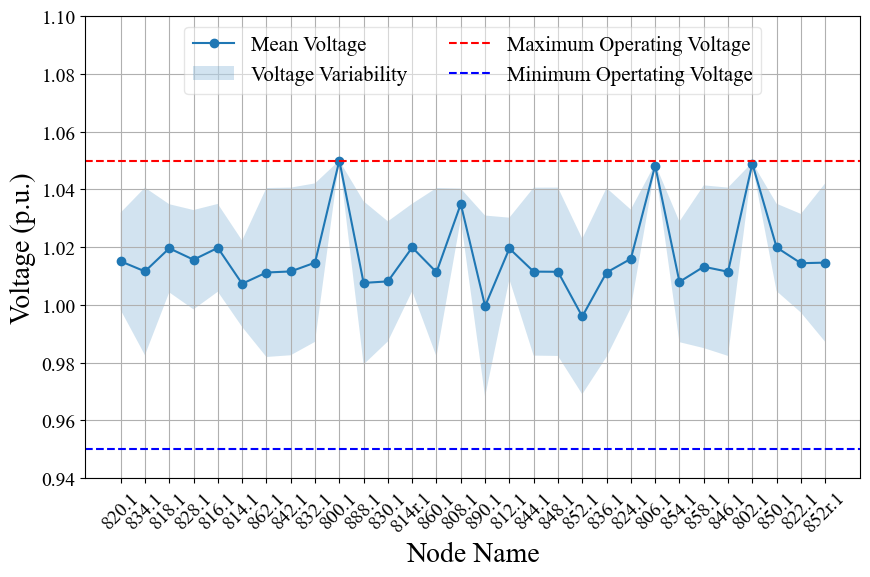}}
\caption{Phase A voltage profile with proposed coordination}
\label{volt_with_coor}
\end{figure}

After implementing the proposed control algorithm for the same scenarios, the resulting voltage (mean and variance) consistently remained within the acceptable range (see Fig. \ref{volt_with_coor}).
The absence of values beyond the designated limits for any of the nodes reflects the algorithm's capability to avert over-voltage (and under-voltage) conditions across the distribution system.
This result highlights its practical utility in maintaining the operational voltage within desirable limits even in presence of considerable amounts of renewable generation. 

The overall process of feeding $\mu$PMU data into ML-based DSSE, and using its outputs to determine the appropriate control action, took less than 1 second. This substantiates the real-time applicability claim of our proposed method (see last paragraph of 
Section \ref{AA}).
Moreover, by learning via system interaction, it circumvents dependency on model parameters.

\section{Conclusion}
The increasing penetration of renewable energy resources in distribution systems necessitates the development and implementation of advanced voltage control strategies. 
Prior approaches for voltage control were often limited by heavy online computational burden and/or requirement of an extensive sensing and communication system.
The proposed DRL-based control strategy relies on estimates obtained from an ML-based state estimator that provides high-speed outputs for real-time unobservable systems.
The results indicate that the proposed infrastructure-light
control effectively maintains voltage across the distribution network, even with the inherent variability introduced by renewable energy resources. 
This research underscores the potential for smart grid technologies to evolve towards more resilient and adaptive management systems in the face of increasing renewable energy utilization.
\bibliographystyle{IEEEtran}
{\footnotesize
\bibliography{References.bib}}

\end{document}